\documentclass[preprint,showkeys,showpacs]{revtex4}
\usepackage{hyperref}
\usepackage{graphicx}
\usepackage{amssymb}
\usepackage{amsmath}

\begin{document}

\title{Exploring Extra Dimensions in Spectroscopy Experiments}
\author{Feng Luo\footnote{
fluo@student.dlut.edu.cn} and Hongya Liu\footnote{
hyliu@dlut.edu.cn}}

\address{Department of Physics, Dalian University of Technology,
Dalian, 116024, P. R. China}



\begin{abstract}
We propose an idea in spectroscopy to search for extra spatial
dimensions as well as to detect the possible deviation from Newton's
inverse-square law at small scale, and we take high-Z hydrogenic
systems and muonic atoms as illustrations. The relevant experiments
might help to explore more than two extra dimensions scenario in
ADD's brane world model and to set constraints for fundamental
parameters such as the size of extra dimensions.
\end{abstract}

\maketitle

\section{Introduction}

The possibility that there exists more than three spatial dimensions
in nature has aroused physicists' interests for many years.
Recently, the studying focus of theories about extra dimensions has
shifted to the \textquotedblleft brane world\textquotedblright\
scenario. In contrast to the original string/superstring theories in
which extra dimensions are assumed to be as small as Planck scale
($\sim 10^{-33}cm$) and hence the detection of these tiny dimensions
is hopeless, extra dimensions in brane world scenario can be large
or even infinite (for a review, see e.g., \cite{Rubakov}), so that
the search for such hidden dimensions becomes much more encouraging.

Among various brane world models, the one proposed by ADD (N.
Arkani-Hamed, S. Dimopoulos and G. Dvali) is especially interesting
\cite{Arkani-Hamed}, \cite {Antoniadis}, \cite{Arkani}, and its
implications in accelerator physics, astrophysics and cosmology have
been widely studied (see, for example, \cite{Hannestad},
\cite{Kanti}, \cite{Luo}). Perhaps the most distinctive character of
this model is it predicts that the familiar Newton's inverse-square
law (ISL) would fail below the size of extra dimensions. The reason
is as follows: to solve the hierarchy problem in particle physics,
that is, the unnatural huge energy gap between Planck mass
$M_{pl}\sim 10^{19}Gev$ and electroweak mass $m_{EW}\sim 10^{3}Gev$,
ADD assume the four dimensional $M_{pl}$ is not a fundamental scale,
but an induced one from the $(4+n)$ dimensional Planck mass
$M_{pl(4+n)}$ through
\begin{equation}
M_{pl}^{2} \sim M_{pl(4+n)}^{2+n}R^{n}, (c=1,\hbar =1), \label{eq1}
\end{equation}
where $n$ and $R$ are the number and radius of the extra dimensions
respectively. For $M_{pl(4+n)}=1Tev$, this equation is equivalent to
\begin{equation}
R\sim 10^{-17+\frac{32}{n}}cm. \label{eq2}
\end{equation}
They further assume $M_{pl(4+n)}$ is around the scale of $m_{EW}$,
then the hierarchy becomes trivial. Considering within brane world
scenario gravitational field (with possible exception of some
hypothetic very weakly coupled fields) is the only field that can
propagate in extra dimensions, a straightforward use of $(4+n)$
dimensional Gauss' law infers that the usual ISL of gravitational
force would change to a much stronger one at small scale as
\begin{equation}
F\propto \frac{1}{r^{n+2}},\text{for }r\ll R, \label{eq3}
\end{equation}
and it recovers to ordinary ISL for $r\gg R$. While the $n=1$ case,
for which $R\sim 10^{12}m$ from Eq.(\ref{eq2}), is excluded from
planetary motion observations, other possibilities of $n$ cannot be
boldly ruled out. Particularly, the $n=2$ case implies
sub-millimeter extra dimensions, while the experimental conditions
for testing Newton's ISL by torsion pendulum method was just about
to be available when ADD's proposal appeared, therefore, many people
have devoted to the search of deviation from ISL as well as extra
dimensions during the past few years \cite{Long}, \cite{Price},
\cite{Adelberger}, \cite{Hoyle}. By now, there is no reported
deviation from ISL and the parameter $M_{pl(4+n)}$ has been
constrained to larger than several $Tev$ by torsion-balance
experiments (\cite{Hoyle}, \cite{Kanti}, including also constraints
from accelerators and astrophysics etc.).

Notice that Eq.(\ref{eq2}) results in smaller $R$ for larger $n$.
For $n=3$, one obtains $R\sim 10^{-7}cm$. Direct measurement of
gravitational force using torsion pendulum at such small scale is
far beyond the reach of current and foreseeable future experimental
ability. (Note, however, that the scales of different extra
dimensions are not guaranteed to be the same, so the deviation from
ISL may also appear in micron range even for $n>2$.) Considering the
limitation of torsion-balance experiments and the importance of
detection of the possible deviation from ISL as well as exploring
extra dimensions, searching for other alternative experimental
methods is worthwhile. It is well know that spectroscopy is one of
the most precise and thoroughly studied fields in physics, so we
wonder whether clues about deviation from ISL and extra dimensions
can be found in such field. The idea is quite simple: since gravity
may become much stronger at small scale, then for small scale
systems, such as atoms, ions and even subatoms, effects of the
original safely neglected gravity may actually be large enough and
able to show itself in spectroscopic spectra. Because of the
precision of spectroscopy, such experiments may help to set
constraints of $M_{pl(4+n)}$ and $R$, or some other model depended
parameters. Moreover, these spectroscopic experiments may be capable
of dealing with the $n=3$ case and detect the possible deviation
from ISL down to nanometer range.

In a previous paper \cite{Liu}, we did some calculations based on
one-proton one-lepton systems, that is, hydrogen atom, muonic
hydrogen etc., and found that the corrections from the strengthened
gravity for the ground state binding energy of these systems are
many orders of magnitude larger than the ones from the exact ISL
gravity. Now we would like further develop our work to high-Z
hydrogenic systems (one electron) and muonic atoms, which are
particular interesting in precision spectroscopy since their
important role in testing quantum electrodynamics (QED) (see, for
example, \cite{Beiersdorfer}, \cite{Beyer}, \cite{Klaft},
\cite{Brown}). The effects of gravity in such systems are much
larger and hence may be more encouraging for our proposal. In this
sense, these systems may be not only suitable to test the Standard
Model of particle physics, but also help to explore scenarios beyond
the Standard Model.

\section{Gravity Corrections in High-Z Hydrogenic Systems}

Even within ADD's framework where gravity becomes much stronger at
small scale, gravity is still far smaller compared to
electromagnetic force in high-Z hydrogenic systems, so we can
conveniently treat it as a perturbation. Also, for estimating
gravity effects from the view of orders of magnitude, we just need
to consider corrections from the leading Schrodinger term.

As an illustration, we will only perform calculations for the ground
state energy level, which is also the simplest case. Clearly,
correction in this energy level is the largest one because of the
smallest Bohr radius, which serves as the mean distance between the
gravitational sources --- the atomic nucleus and the electron.

The first-order correction of the ground state energy level is
written as
\begin{equation}
\Delta E=\int_{0}^{\infty} \int_{0}^{2 \pi} \int_{0}^{\pi}
\Psi_{100}^{\ast}  \hat{V} (r) \Psi_{100}r^{2}\sin\theta d\theta
d\phi dr, \label{eq4}
\end{equation}
where $\Psi _{100}=\pi
^{-\frac{1}{2}}a^{-\frac{3}{2}}e^{-\frac{r}{a}}$ is the ground state
wave function of the high-Z hydrogenic system, $a=\frac{\hbar
^{2}}{me^{2}Z}$ is the first Bohr radius,
$m=\frac{m_{l}{M}}{m_{l}+M}$ is the reduced mass, $m_{l}$ and $M$
are the masses of electron and atomic nucleus respectively. The
gravitational potential is given as
\begin{equation}
\hat{V} (r)=\left\{
\begin{array}{cc}
-\frac{G_{(4+n)}m_{l}M}{r^{n+1}}, & r\ll R \\
-\frac{G_{4}m_{l}M}{r}(1+\alpha e^{-\frac{r}{\lambda }}), & r\sim R
\\
-\frac{G_{4}m_{l}M}{r}, & r\gg R%
\end{array}%
\right. , \label{eq5}
\end{equation}
where the $(4+n)$ dimensional Newton's constant is $G_{(4+n)}\sim
R^{n}G_{4}$, and the four dimensional Newton's constant is
$G_{4}=M_{pl}^{-2}$. The second line of Eq.(\ref{eq5}) takes the
Yukawa type, in which the parameters $\alpha \sim n$ and $\lambda
\sim R$. In above expressions, we have neglected constants of order
unity which depends on the specific compactification forms of extra
dimensions, and we have assumed that the sizes of all the extra
dimensions are the same. More detail expressions can be seen in
\cite{Adelberger}, \cite{Liu}.

From Eq.(\ref{eq4}) and the first line of Eq.(\ref{eq5}), one can
notice that the integral diverges as $r\rightarrow 0$ for $n\geq2$.
Considering the atomic nucleus is not point like, we introduces a
safe cutoff value $r_{m}$ with the atomic nucleus size for the lower
limit of the integral. A convenient expression of $r_{m}$ is
$r_{m}=r_{0}A^{\frac{1}{3}}$, where $A$ is the mass number of the
atomic nucleus and $r_{0}$ is of size $\sim10^{-13}cm$. Also note
that such cutoff leads to a relatively conservative estimation of
the gravity effects.

Neglecting coefficients of order unity, the above equations give
\begin{equation}
\Delta E\sim \left\{
\begin{array}{cc}
-G_{4}m_{l}Ma^{-1}, &n=0\\
-M_{pl(4+n)}^{-3}m_{l}Ma^{-2}, &n=1\\
-M_{pl(4+n)}^{-4}m_{l}Ma^{-3}, &n=2\\
-M_{pl(4+n)}^{-5}m_{l}Ma^{-3}r_{m}^{-1}, &n=3\\
-M_{pl(4+n)}^{-6}m_{l}Ma^{-3}r_{m}^{-2}, &n=4\\
\end{array}
\right. . \label{eq6}
\end{equation}
Since $r_{m}\sim10^{-12}cm$ and $M_{pl(4+n)}^{-1}\sim 10^{-17}cm$
for $M_{pl(4+n)}\sim1Tev$, then for $n\geq4$, the corrections are
much smaller than the ones for $n=2$ and $n=3$, which are the cases
of our interest.

In fact, for $n=2$ and $n=3$, $\Delta E \propto AZ^{3}$ and $\Delta
E \propto A^{\frac{2}{3}}Z^{3}$ respectively. Therefore, it is clear
that the corrections for high-Z hydrogenic systems are much larger
compared to the ones for hydrogen atom. Convert $\Delta E$ to
corrections in terms of frequency $\Delta \nu$ through $\Delta E=h
\Delta \nu$. For a specific system, say $^{207}$Pb$^{81+}$, $\Delta
\nu \sim 10^{0}Hz$ for $n=2$ and $10^{-5}Hz$ for $n=3$, while the
corresponding corrections in hydrogen atom are $10^{-8}Hz$ and
$10^{-13}Hz$ respectively. Also note that the exact ISL, that is,
the $n=0$ case gives a correction as small as $\Delta \nu \sim
10^{-24}Hz$ for hydrogen atom.

The above calculations also suit to muonic atoms (one muon in the
atoms or ions, not confined to hydrogen-like), since the ground
state muonic orbit is far interior compare to the first electronic
Bohr orbit and hence the muon sees approximately a pure Coulomb
potential from the atomic nucleus. In this case, what we need to do
is just assigning $m_{l}$ in Eq.(\ref{eq6}) the mass of muon, which
is about $200$ times larger than electron. Notice $m_{l}$ also
appears in the expression of $a$, such substitution further results
in several orders of magnitude increase of $\Delta \nu$. For
$^{207}$Pb muonic atom, $\Delta \nu \sim 10^{9}Hz$ for $n=2$ and
$10^{4}Hz$ for $n=3$. For a medium-Z muonic atom, e.g., $^{40}$Ca
muonic atom, $\Delta \nu \sim 10^{6}Hz$ for $n=2$ and $10^{2}Hz$ for
$n=3$.

\begin{figure}[!t]
\centering
\includegraphics[width=4.0in]{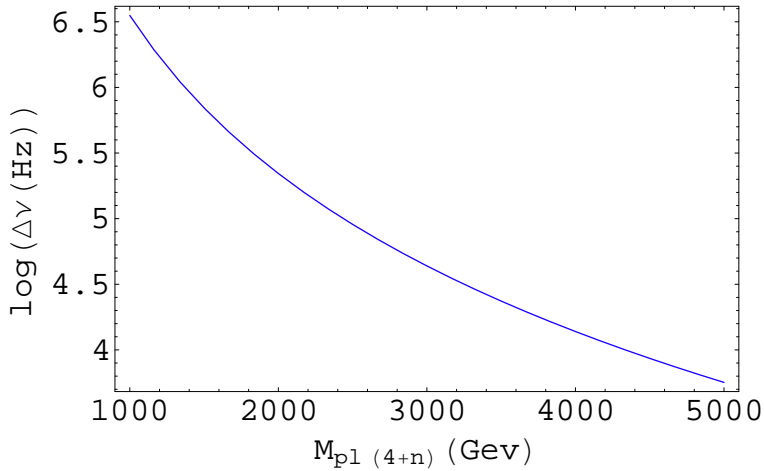}
\caption{Dependence of $\Delta \nu$ to $M_{pl(4+n)}$ for $^{40}$Ca
muonic atom, $n=2$.} \label{F1}
\end{figure}

\begin{figure}[!t]
\centering
\includegraphics[width=4.0in]{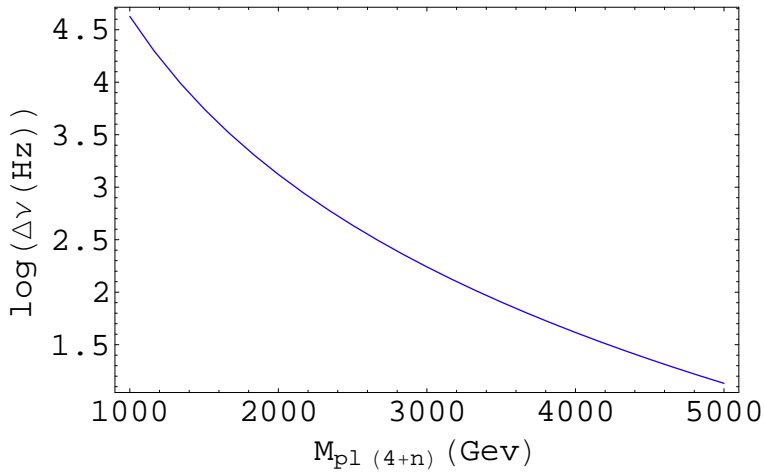}
\caption{Dependence of $\Delta \nu$ to $M_{pl(4+n)}$ for $^{207}$Pb
muonic atom, $n=3$.} \label{F2}
\end{figure}

The sensitivity of $\Delta \nu$ to $M_{pl(4+n)}$ and the
corresponding $R$ (notice Eq.(\ref{eq1})) is easily seen from
Eq.(\ref{eq6}), since $\Delta \nu \propto M_{pl(4+n)}^{-4}$ and
$\Delta \nu \propto M_{pl(4+n)}^{-5}$ for $n=2$ and $n=3$
respectively. Fig.~\ref{F1} show the $^{40}$Ca muonic atom case for
$n=2$, and Fig.~\ref{F2} show the $^{207}$Pb muonic atom case for
$n=3$.

\section{Conclusions and Discussions}

Considering the above large corrections from gravitational effects,
it appears spectroscopic experiments using high-Z hydrogenic systems
and medium-Z or high-Z muonic atoms may serve as new methods to
detect the possible deviation of Newton's ISL as well as to search
for extra dimensions. The high precision X-ray and laser
spectroscopy make this proposal interesting, since the relevant
experiments may not only be able to explore the $n=2$ case in ADD's
brane world model, but also make the study of $n=3$ case and
nanometer range deviation possible. Also, the strong dependence of
the corrections to fundamental parameters of the model, e.g., the
scales of extra dimensions, may help to further constrain such
parameters.

We note, however, great efforts may have to be made before the
realization of our proposal. Theoretically, the uncertainties of the
nuclear structure effects, the higher-order terms neglected in
calculation of other effects etc., may submerge the effects of
gravity. Experimentally, to achieve enough accuracy to show the
strengthened gravity effects is also challenging (for a review about
experimental accuracy for Lyman-$\alpha$ transitions in high-Z
hydrogenic systems, see \cite{NPL}).

Anyway, although difficulties exist, it is fair to say that in
precision spectroscopy, similar ideas may find their way in many
other systems, e.g., subatomic and sub-nuclear systems, and by
various experiments, e.g., experiments involving hyperfine
structures. In a word, precision spectroscopy may play an active
role in the development of the most exotic theories of particle
physics and cosmology, and it may really help to explore those most
exotic mysteries like extra dimensions.

\section{Acknowledgements}

This work was supported by NSF (10573003) and NBRP (2003CB716300) of
P. R. China.

\end{document}